\documentclass[
twocolumn,
superscriptaddress,
 amsmath,amssymb,
 aps,
 prl,
]{revtex4}
\usepackage{dcolumn}
\usepackage{bm}
\usepackage{graphicx}
\usepackage{comment}

\begin{document}

\title{Universal survival probability for a $d$-dimensional run-and-tumble particle}
\author{Francesco Mori}
\affiliation{LPTMS, CNRS, Univ.  Paris-Sud,  Universit\'e Paris-Saclay,  91405 Orsay,  France}
\author{Pierre Le Doussal}
\affiliation{Laboratoire de Physique de l'Ecole Normale Sup\'erieure, PSL University, CNRS, Sorbonne Universit\'es, 24 rue Lhomond, 75231 Paris, France}
\author{Satya N. Majumdar}
\affiliation{LPTMS, CNRS, Univ.  Paris-Sud,  Universit\'e Paris-Saclay,  91405 Orsay,  France}
\author{Gr\'egory Schehr}
\affiliation{LPTMS, CNRS, Univ.  Paris-Sud,  Universit\'e Paris-Saclay,  91405 Orsay,  France}

\date{\today}

\begin{abstract}

We consider an active run-and-tumble particle (RTP) in $d$ dimensions and compute exactly the probability $S(t)$ that 
the $x$-component of the position of the RTP does not change sign up to time $t$. When the tumblings occur at a constant rate, we show that $S(t)$ is independent of $d$ for any finite time $t$ (and not just for large $t$), as 
a consequence of the celebrated Sparre Andersen theorem for discrete-time random walks in one dimension. 
Moreover, we show that this universal result 
holds for a much wider class of RTP models in which the speed $v$ of the particle after each tumbling is random,  
drawn from an arbitrary probability distribution. We further demonstrate, as a consequence, the universality of the record statistics in the RTP problem.

\end{abstract}

\maketitle

\newpage
\begin{figure}[t]
\includegraphics[scale=0.24]{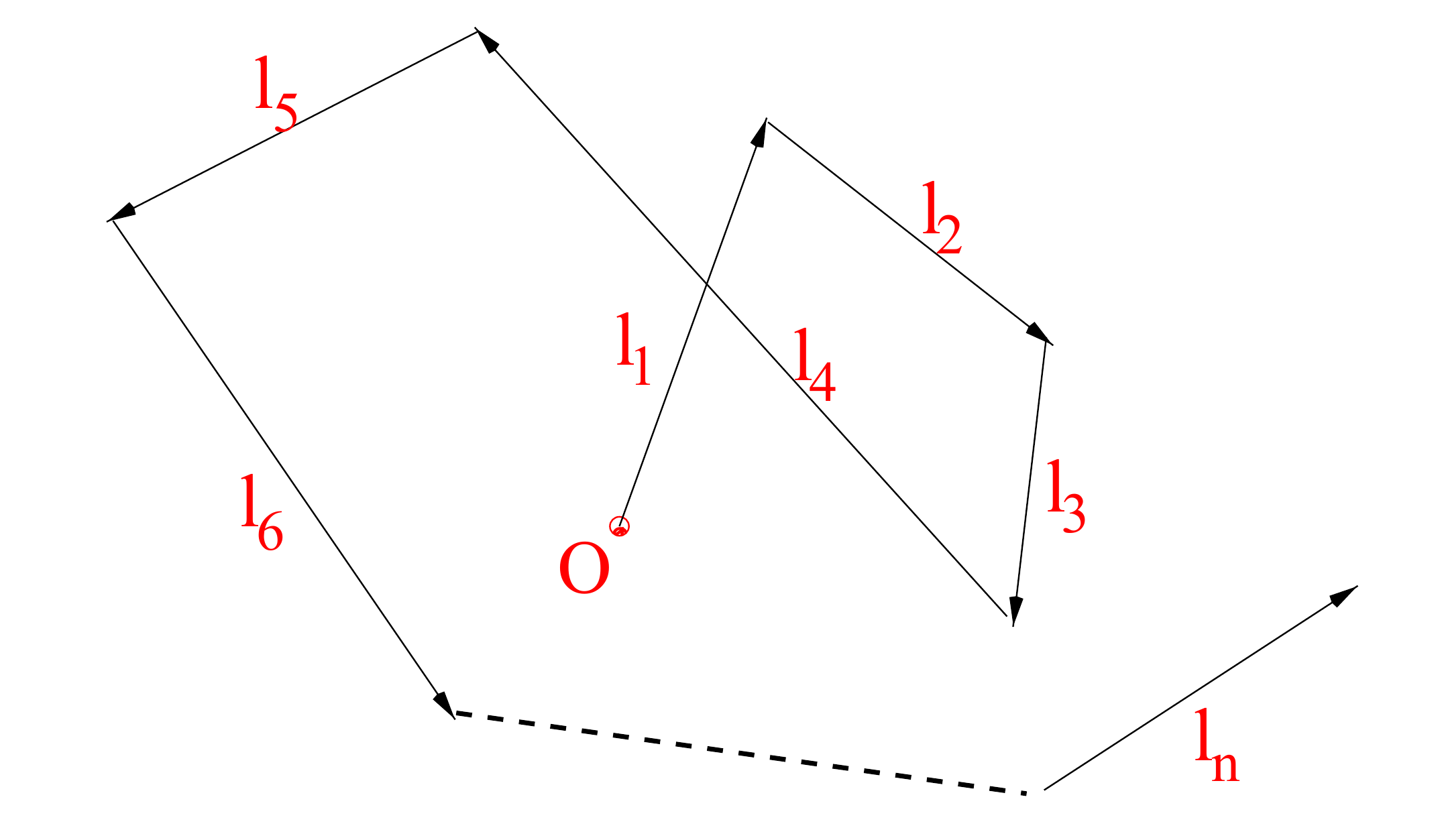} 
\caption{\label{fig:rtm} Typical trajectory of a RTP in two dimensions. The particle starts at the origin $O$, chooses a random direction and moves ballistically in that direction for a distance $l_1=v_0\, \tau_1\, ,$ where $v_0$ is constant and $\tau_1$ is a random time drawn from the exponential distribution $p(\tau)= \gamma\, e^{-\gamma \tau}$. At the end of this first flight, the particle tumbles instantaneously and chooses a new random direction and again moves ballistically a distance $l_2= v_0 \tau_2$ with $\tau_2$ drawn independently from the same $p(\tau)$. Then the particle tumbles again and so on.}
\end{figure}

The first time $t_{\rm f}$ at which a stochastic process reaches a fixed target level is a fundamental 
observable with many applications. Statistics of $t_{\rm f}$ plays a crucial role in various situations, including e.g., the 
encounter of two molecules in a chemical reaction \cite{HTB90}, the capture of a prey in a hunting scenario \cite{BLMV11}, 
or the escape of a comet from the solar system \cite{Ham61, BF_2005}. In the context of finance, agents often 
use limit orders to buy/sell a stock only when its price is below/above a target value. Thus, it is important to estimate 
if and when that target value will be reached and this question has been intensively studied during 
decades (for recent reviews see \cite{BLMV11,Redner_book, SM_review, Persistence_review, fp_book_2014,Masoliver_book}).  
Due to the ubiquity of these problems, novel applications are constantly being identified, raising in turn new 
challenging questions.

In recent years, tremendous efforts have been devoted to the study of statistical fluctuations in active matter systems 
\cite{Marchetti13, Bechinger16, Ram17,Sch03}. 
In contrast to a passive matter such as a Brownian motion (BM), whose dynamics is driven by thermal fluctuations of the 
environment, this class of active non-equilibrium systems is characterized by self-propelled motility based on continuous 
consumption of energy from the environment. For example, models of active matter have been used to describe vibrating 
granular matter \cite{WW_2017}, active gels \cite{R_2010,NVG2019}, bacteria \cite{Berg_book, Cates_bacteria} or collective motion 
of ``animals'' \cite{R_2010,V_1995,HB_2004, VZ12}. In this context, one of the most studied model is the run-and-tumble 
particle (RTP) \cite{TC_2008, CT_2015}, also known as ``persistent random walk'' \cite{Weiss_2002,ML_2017}. 
In the simplest version of the model, an RTP performs a ballistic motion along a certain 
direction at a constant speed $v_0\ge 0$ (``run'') during a 
certain ``time of flight'' $\tau$. Following this run, it ``tumbles'', i.e., chooses a new direction 
uniformly at random and then performs a new run along this direction again with speed $v_0$
during a random time $\tau$ and so on (see Fig.~\ref{fig:rtm}). 
Typically these tumblings occur with constant rate $\gamma$, i.e. the $\tau$'s of different runs are independently distributed via 
exponential distribution $p(\tau) = \gamma e^{-\gamma \tau}$, though other distributions will also be considered later. 
Despite its 
simplicity, this RTP model exhibits complex interesting features such as clustering at boundaries \cite{Bechinger16}, 
non-Boltzmann distribution in the steady state in the presence of a confining 
potential~\cite{TC_2008, Dhar_18,Sevilla_19,MBE2019,3states_19}, motility-induced phase 
separation \cite{CT_2015}, jamming~\cite{SEB2016} etc. 
Variants of the RTP model where the speed $v\ge 0$ of the particle is renewed after each 
tumbling by drawing it from a probability density function (PDF) $W(v)$~\cite{GM_2019,footnote_andrea} or where the RTP 
undergoes random resetting to its initial position at a constant rate~\cite{EM_2018,M2019} have also been studied.

In the $d=1$ case, the first-passage properties of the RTP model and of its variants have been widely 
studied~\cite{Weiss_2002,artuso14,Malakar_2018,DM_2018,LDM_2019}. Several recent studies investigated the 
survival probability of an RTP in $d=1$, both in the absence and in the presence of a confining potential/wall \cite{Malakar_2018,DM_2018,LDM_2019,ADP_2014,Dhar_18}. The $d=1$ case is analytically tractable because the velocity has only two possible directions $\pm v_0$, which simplifies the problem in $d=1$. However, in 
$d>1$, the first-passage problems become much more difficult because the orientation of the velocity 
is a continuous variable. Consequently, exact results are difficult to obtain in $d \geq 2$, though approximation
schemes have been developed recently for the mean first-passage time in a confined geometry~\cite{RBV16}.

\begin{figure}[t]
\includegraphics[width = 0.75\linewidth]{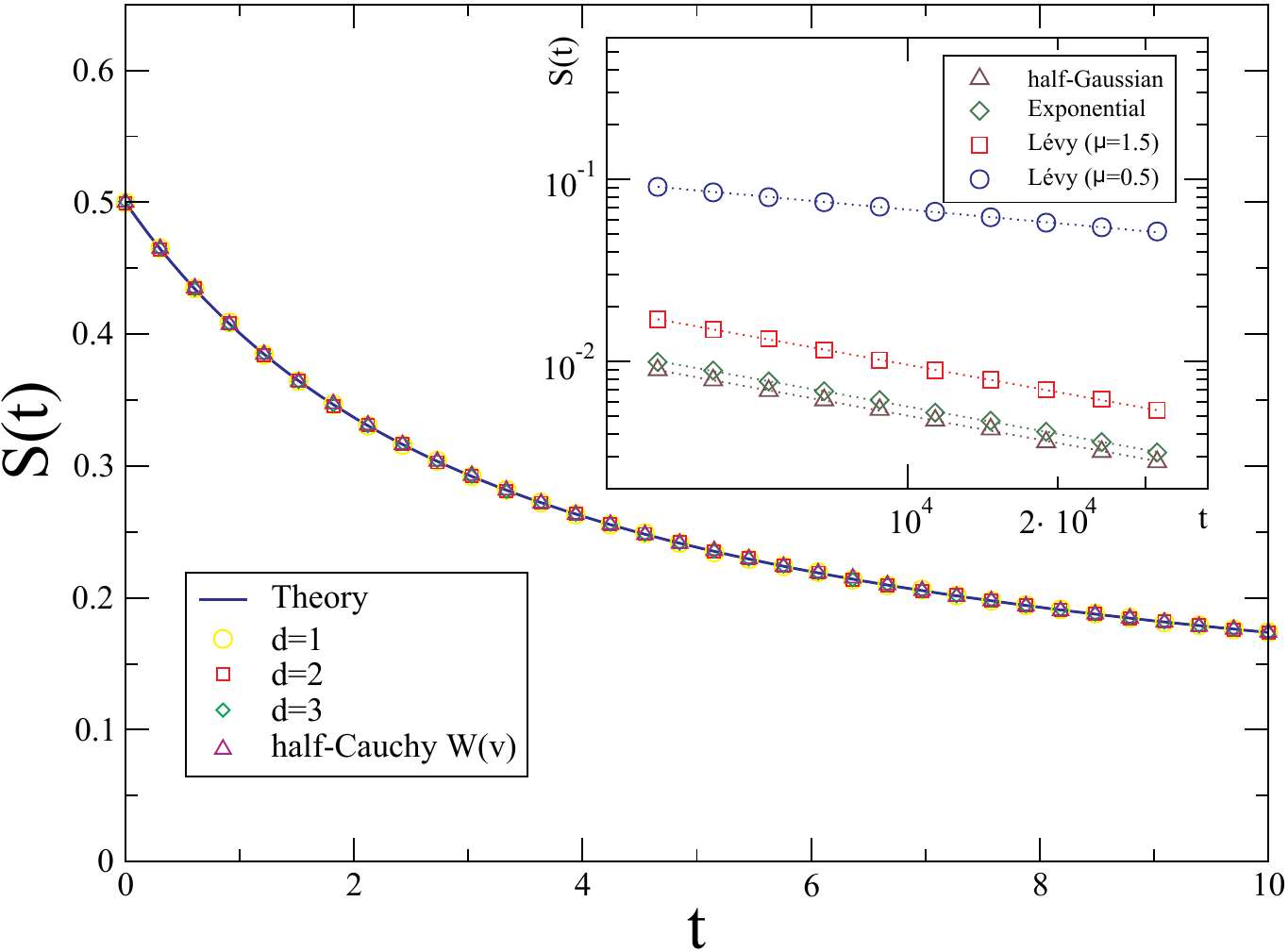} 
\caption{\label{fig:s} Survival probability $S(t)$ as a function of time $t$, for $\gamma=1$. 
The continuous blue line corresponds to the exact result in Eq. (\ref{eq:s}). The symbols correspond to
simulations with the choices $d=1,2,3$, $W(v) = \delta(v-1)$ and $p(\tau) = e^{-\tau}$ and one case where
$d=2$, $W(v) = 2/(\pi (1+v^2))$ with $v>0$ (half-Cauchy) and $p(\tau) = e^{-\tau}$. They all fall on the analytical blue line for all $t$. \textbf{Inset:} Numerical computation of $S(t)$ in $d=2$ for different distributions $p(\tau)$: (i) half-Gaussian, (ii) exponential, (iii)  asymmetric L\'evy distribution with L\'evy index $\mu = 3/2$ and (iv) asymmetric L\'evy with $\mu =1/2$. In all these cases, $S(t) \sim t^{-\theta}$ for large $t$ with $\theta = \frac{1}{2}$ in cases (i)-(iii) and $\theta = \mu/2 = 1/4$ for case (iv) corresponding to $\mu=1/2$.}
\end{figure}

In this Letter we consider an RTP in $d$-dimensions, starting from the origin with a random velocity, and compute exactly the 
probability $S(t)$ that the $x$-component of the RTP does not change sign up to time~$t$. It is useful to view $S(t)$ as 
the ``survival probability'' of the RTP in the presence of an absorbing hyperplane passing through the origin and 
perpendicular to the $x$-axis. For a passive particle executing Brownian motion (BM), it is clear that $S(t)$ is 
independent of $d$, since each component of the displacement performs an independent one-dimensional BM~\cite{footnote_BM}. In contrast, for an RTP in $d$ dimensions, the different spatial components are coupled (see Fig. 
\ref{fig:rtm}) and, consequently one may expect that $S(t)$ would depend on the dimension $d$ and the speed $v_0$. 
Performing first simulations in $d=1,2,3$ (see Fig. \ref{fig:s}) we found, rather amazingly, that $S(t)$ is completely 
independent of both $d$ and $v_0$, at {\it any finite time} $t$ (and not just at large times only)\,!

The principal goal of this Letter is to understand and prove this remarkably universal result valid even at finite $t$.
We compute $S(t)$ exactly for all $t$ in arbitrary dimension $d$, and demonstrate that it is 
indeed independent of the dimension $d$ and speed $v_0$ for any time $t$ and is given by a simple formula 
\begin{equation}
\label{eq:s}
S(t)=\frac{1}{2}e^{-\gamma t /2} \left( I_0 \left(\gamma t /2 \right) + I_1 \left( \gamma t /2\right) \right) \;,
\end{equation}
where $I_0(z)$ and $I_1(z)$ are modified Bessel functions. 
When $t\rightarrow0$, $S(t)$ goes to the limiting value $1/2$, 
which is just the probability that the $x$-component of the initial direction is positive. 
On the other hand, at late times it decays as $S(t) \sim 1/\sqrt{\pi \gamma t}$. 
By mapping our $d$-dimensional process to an effective $1d$-process
we show below that the universality of this result (\ref{eq:s}) is inherited from the 
universality of the Sparre Andersen (SA) theorem \cite{SA_54} for the survival probability of a 
one-dimensional discrete-time random walk. In the special case $d=1$, as a bonus, we recover here
using a completely different method, the result 
in Eq. (\ref{eq:s}) obtained in 
previous works \cite{Weiss_2002,Malakar_2018,LDM_2019} via Fokker-Planck approaches. 
In Fig. \ref{fig:s}, we compare our formula for $S(t)$ in (\ref{eq:s}) with  
numerical simulations for $d=1,2$ and $3$, finding an excellent agreement at all $t$.
Furthermore, this universal result (\ref{eq:s}) also holds for a broader class of RTP 
models where the speed $v$, and not just the direction, is also renewed afresh after each 
tumbling, chosen each time independently from the PDF $W(v)$, with $v\in [0,\infty)$. 
The standard RTP model corresponds to the choice $W(v)=\delta(v-v_0)$ but
this also includes fat tailed PDF
$W(v)$ such as the half-Cauchy distribution: $W(v)= 2/[\pi\,(1+v^2)]$ ($v\ge 0$), as shown in Fig. \ref{fig:s} .
Our main result thus states that for the most common RTP model with exponentially distributed time of flights
$p(\tau) = \gamma e^{-\gamma \tau}$, 
the survival probability $S(t)$ at all $t$ 
is not only independent of the dimension $d$, but also on the velocity distribution $W(v)$ and  
is given by Eq.~(\ref{eq:s}). We further show that this universal behavior ceases to hold if the distribution of the 
$\tau$'s is not an exponential. In fact, if $p(\tau)$ has a well defined first moment then one still has $S(t) \propto 
t^{-1/2}$ at large times but $S(t)$ is not universal for finite time $t$. Finally, for very fat tailed distribution such 
that the first moment is not defined, e.g. for $p(\tau) \propto \tau^{-1-\mu}$ for large $\tau$ with $\mu<1$ (in the $1d$ case this corresponds to L\'evy walks, see e.g.  \cite{metzler}), then $S(t) 
\propto t^{-\mu/2}$ as $t \to \infty$ but again the finite $t$ behavior of $S(t)$ is not universal.

Interestingly, the SA theorem was also used recently~\cite{artuso14} to compute first-passage statistics
in a variant of the one dimensional RTP model. In this ``wait-then-jump model'' 
the particle waits a random time during tumbling and 
then jumps instantaneously to a new position. Combining the SA theorem with additional combinatorial arguments,
the authors of Ref.~\cite{artuso14} derived nice results for general jump distributions in their `wait-then-jump' model. 
Unfortunately, their clever method can not be adapted to compute the survival probability in the standard RTP model
considered here, where the trajectory of the particle is continuous in time. 
In fact our method turns out to be more general: it not only provides an exact solution for the standard RTP problem
in $d$-dimensions and its generalization to RTP's with an arbitrary speed distribution $W(v)$, but also recovers the results of Ref.~\cite{artuso14} by a simpler non-combinatorial
method (see \cite{supmat} for details).

To sketch the derivation of our main result in Eq.~(\ref{eq:s}), we consider a typical trajectory of an RTP in $d$-dimensions, 
starting at the origin $O$ at $t=0$ (see Fig. \ref{fig:rtm}). For simplicity, we start with the case $W(v)=\delta(v-v_0)$.  
Note that in a fixed time window $t$, the number of tumblings $n$ undergone 
by the particle is a random variable and varies from trajectory to trajectory. We will
count the starting point $O$ as a tumbling, which implies $n\geq 1$. The time 
$\tau$ between two tumblings is drawn from $p(\tau)=\gamma e^{-\gamma \tau}$ independently after
each tumbling 
and we denote the time interval after the $i^{\rm th}$ tumbling as $\tau_i$. Note that the duration
 $\tau_n$ of the last interval  travelled by the particle before the final time $t$ is yet to be completed. 
Hence, the probability of no tumbling during that time interval is $\int_{\tau_n}^\infty p(\tau) d\tau = e^{-\gamma\tau_n}$. 
Thus, the joint distribution of the time intervals $\{\tau_i\}=\{\tau_1,\tau_2,...,\tau_n\}$ and the
number of tumblings $n$, for a fixed duration $t$, is given by
\begin{equation}
P\left(\{\tau_i\}, n|t\right)= 
\left[\prod_{i=1}^{n-1} \gamma\, e^{-\gamma\, \tau_i}\right]\, 
e^{-\gamma\, \tau_n}\, \delta\left( \sum_{i=1}^n \tau_i-t\right),
\label{joint_tau_n.1}
\end{equation}
where the $\delta$ function enforces the constraint that the total time is $t$. 
Let $\{l_i\}=\{l_1,l_2,...,l_n\}$ denote the straight distances travelled by the particle up to time $t$ 
(see Fig. \ref{fig:rtm}). Clearly $l_i=v_0 \tau_i$ and $l_i\geq 0 $ for all $i$. Consequently, using Eq. (\ref{joint_tau_n.1}), 
the joint distribution of $\{l_i\}$ and the number of tumblings $n$ is given by
\begin{equation}
P\left(\{l_i\}, n|t\right)= \frac{1}{\gamma}\,
\left[\prod_{i=1}^{n} \frac{\gamma}{v_0}\, e^{-\gamma\, l_i/v_0}\right]\,
\delta\left( t- \sum_{i=1}^n \frac{l_i}{v_0}\right).
\label{joint_l_n.1}
\end{equation}
We now want to write the joint distribution of the $x$-components of these random vectors $\{\vec l_i\}$
with given norms $\{l_i\}$. To proceed, we consider a random vector $\vec l$ in $d$ dimensions whose norm 
$l= |\vec l|$ is fixed and whose direction is uniformly distributed. 
Let $x$ denote the $x$-component of this random vector $\vec l$. The distribution of this $x$-component {\it given} 
the fixed norm $l$ is $P_d(x|l)= (1/l) f_d (x/l)$, with (for derivation see \cite{supmat})
\begin{equation}
f_d(z)= \frac{\Gamma(d/2)}{\sqrt{\pi}\, \Gamma((d-1)/2)}\, (1-z^2)^{(d-3)/2}\, \theta(1-|z|)\,,
\label{fdz.1}
\end{equation}  
where $\Gamma(y)$ is the Gamma function and $\theta(y)$ is the Heaviside step function: $\theta(y)=1$ if $y\ge 0$ and 
$\theta(y)=0$ if
$y<0$. We denote as $x_i$ the $x$-component of the vector $\vec{l_i}$. Since at each tumbling the new 
direction is drawn independently, the joint probability distribution of the $x$-components, 
given the distances $\{l_i\}$ factorises as 
$P(\{x_i\}|\{l_i\})=\prod_{i=1}^n (1/l_i) f_d (x_i /l_i)$. Using this result and Eq. (\ref{joint_l_n.1}), 
we can now write the joint probability distribution of $\{x_i\}$, $\{l_i\}$ and $n$ as
\begin{equation}
\begin{split}
P(\{x_i\},\{l_i\},n|t)=P(\{x_i\}|\{l_i\})P(\{l_i\},n|t)\\=
\frac{1}{\gamma} \left[ \prod_{i=1}^n \frac{1}{l_i}
f_d\left(\frac{x_i}{l_i}\right)\, \frac{\gamma}{v_0}\, e^{-\gamma\, l_i/v_0}\right]\,
\delta\left( t- \sum_{i=1}^n \frac{l_i}{v_0}\right).
\label{full_joint.1}
\end{split}
\end{equation} 

By integrating over the $\{l_i\}$ variables, we obtain the joint distribution 
of the $x_i$'s and $n$, $P(\{x_i\},n|t)$. Due to the presence of the delta-function 
in (\ref{full_joint.1}), it is convenient to compute its Laplace transform with respect to (w.r.t) $t$. 
After integrating over the $l_i$'s, we obtain (see \cite{supmat}) 
\begin{equation}
\int_0^{\infty} dt\, e^{-s\,t}\, P\left(\{x_i\}, n|t\right) = 
\frac{1}{\gamma}\, \left(\frac{\gamma}{\gamma+s}\right)^n \prod_{i=1}^n {\tilde p}_s(x_i),
\label{lt_joint.1}
\end{equation}
where we have defined
\begin{equation}
{\tilde p}_s(x)= \int_0^{\infty} \frac{dl}{l} f_d\left(\frac{x}{l}\right)\, 
\frac{(\gamma+s)}{v_0}\, e^{-(\gamma+s)\,l/v_0}\, .
\label{ptilde.1}
\end{equation}
One can easily check that ${\tilde p}_s(x)$ is non-negative and normalised to unity (see \cite{supmat}): it can thus be
interpreted as a PDF, parametrized by $s$, $d$, $\gamma$ and $v_0$. 
Moreover, due to the symmetry of $f_d(z) = f_d(-z)$, ${\tilde p}_s(x)$ is also symmetric, i.e., 
${\tilde p}_s(x)={\tilde p}_s(-x)$. While the PDF ${\tilde p}_s(x)$ in Eq. (\ref{ptilde.1}) can be computed 
explicitly for arbitrary $d$, we will show that its precise expression is not relevant for our purpose. 
All that matters for our purpose is
that it is continuous and symmetric in $x$. By performing a formal inversion of the Laplace transform 
in Eq. (\ref{lt_joint.1}), we derive the joint distribution of the $x_i$'s and $n$, given $t$,
\begin{equation}
P\left(\{x_i\}, n|t\right)=  
\int \frac{ds}{2\pi\,i} e^{s\, t}\, 
\frac{1}{\gamma}\, \left(\frac{\gamma}{\gamma+s}\right)^n \prod_{i=1}^n {\tilde p}_s(x_i) \;,
\label{Pxn.1}
\end{equation}
where the integral is over the Bromwich contour (imaginary axis in this case) in the complex $s$ plane. We see from Eq. (\ref{Pxn.1}) 
that the $d$-dimensional RTP (see Fig. \ref{fig:rtm}), when projected in the $x$-direction, constitutes an effective 
one-dimensional random walk (RW)
where the increments $x_i$'s are now correlated in a nontrivial way. Our goal is now to compute the survival 
probability $S(t)$ for this RW, starting from $x_0=0$.

To proceed, we notice that the survival probability $S(t)$ of this $x$-component process up to time $t$ is, by definition, the 
probability of the event that the successive sums $x_1$, $x_1+x_2$, $\ldots$, $x_1+x_2+\ldots x_n$ are all positive. Here, the 
number of steps $n$ of the RW, i.e. the number of tumblings in the initial RTP problem, in the fixed time interval $[0,t]$ is 
itself a random number. Hence, to compute $S(t)$ we need to sum over all possible values of $n \geq 1$. This yields

\begin{equation}
S(t) = \sum_{n=1}^{\infty} \int_{-\infty}^{\infty} dx_1\ldots \int_{-\infty}^{\infty} dx_n\,
\Theta_n(\{x_i\})\, P\left(\{x_i\}, n|t\right)\, ,
\label{St.1}
\end{equation}
where we used the notation $\Theta_n(\{x_i\})=\theta(x_1)\theta(x_1+x_2)\ldots
\theta(x_1+x_2+\ldots+x_n)$  
to constrain the partial sums to be positive. By inserting the expression of 
$P\left(\{x_i\}, n|t\right)$ given in (\ref{Pxn.1}) into Eq. (\ref{St.1}) we obtain
\begin{eqnarray}
S(t) = \int \frac{ds}{2\pi\,i} e^{s\, t}\,
\frac{1}{\gamma}\, \sum_{n=1}^{\infty} \left(\frac{\gamma}{\gamma+s}\right)^n\, q_n \, ,
\label{St.2}
\end{eqnarray}
where we have defined the multiple integral
\begin{equation}
q_n= \int_{-\infty}^{\infty} dx_1\ldots \int_{-\infty}^{\infty} dx_n\,
\Theta_n(\{x_i\})\, \prod_{i=1}^n {\tilde p}_s(x_i)\, .
\label{qn.1}
\end{equation}
In fact, $q_n$ in Eq. (\ref{qn.1}) has a very simple and nice interpretation. Consider a discrete-time 
continuous-space random walk starting at the origin $0$ in one dimension. At each step $k\geq 1$, the position
of the random walker $X_k$ jumps by a random distance $x_k$ drawn, 
independently at each step, from the continuous and symmetric PDF ${\tilde p_s}(x)$, i.e. $X_k = X_{k-1} + x_k$, 
starting from $X_0=0$. Then, $q_n$ in (\ref{qn.1}) just denotes the probability that the walker stays on the 
positive side up to step $n$. Since the jump distribution ${\tilde p}_s(x)$ is continuous and symmetric, 
we can use the Sparre Andersen theorem \cite{SA_54} which states that $q_n$ is universal, i.e. 
independent of ${\tilde p}_s(x)$, and simply given by $q_n= {2n \choose n}\, 2^{-2n}$ for $n\geq 0$. 
Note that this formula is independent of the jump distribution for all $n$, and not just asymptotically for large $n$. 
The generating function of $q_n$ is thus also universal and given by
\begin{eqnarray}\label{eq:SA}
\sum_{n=0}^{\infty} q_n\, z^n = \frac{1}{\sqrt{1-z}} \;.
\end{eqnarray}
This formula has been used recently in several statistical physics problems~\cite{Louven_review,Persistence_review}, 
in particular in the context of record statistics \cite{Ziff_Satya,PLD_2009,MSW2012,GMS2016,Record_review} (see also below and in \cite{supmat} for the record statistics in the RTP problem).
Here we use this result (\ref{eq:SA}) choosing $z = \gamma/(\gamma + s)$ in Eq. (\ref{St.2}),  
taking care of the fact that the sum in Eq. (\ref{St.2}) does not include the $n=0$ term. 
This leads to our amazingly universal result  
\begin{equation}
S(t)= \int \frac{ds}{2\pi\,i} e^{s\, t}\,
\frac{1}{\gamma}\ \left[\sqrt{\frac{\gamma+s}{s}}-1\right]\\ .
\label{sa.3}
\end{equation}
This result is evidently independent of the dimension $d$ and the speed $v_0$. 
The dimensional dependence appears in Eq.~(\ref{St.2}) through the PDF ${\tilde p}_s(x)$ which
however disappears as a consequence of the SA theorem. The Laplace inversion in Eq. (\ref{sa.3}) can be 
exactly done and we obtain the explicit expression for $S(t)$ presented in Eq. (\ref{eq:s}). 
Let us emphasize, once more, that the result (\ref{eq:s}) is valid at all times $t$ and in any dimension $d$. 



In fact, the result (\ref{eq:s}) turns out to be valid for a much broader class of $d$-dimensional RTP models where 
the speed during a flight
is itself a random variable, drawn from a generic speed distribution $W(v)$ -- 
while the time of flights are still exponentially distributed, i.e. $p(\tau) = \gamma \, e^{-\gamma \tau}$.  
For a general $W(v)$, all the steps of our calculation leading to $S(t)$ in (\ref{St.2}) and (\ref{qn.1}) go through, 
except that ${\tilde p}_s(x)$ 
in Eq.~(\ref{ptilde.1}) gets modified to \cite{supmat}
\begin{equation}
{\tilde p}_s(x)= \int_0^{\infty} \frac{dl}{l} f_d\left(\frac{x}{l}\right)\,
\int_0^{\infty} dv\, W(v)\, \frac{(\gamma+s)}{v}\, e^{-(\gamma+s)\,l/v} \;,
\label{ptilde.2}
\end{equation}
which is normalized to unity and is both continuous and symmetric. 
Using the SA theorem, we then conclude that 
$S(t)$ is again independent of the precise form of $\tilde p_s(x)$ and is given by the same universal formula (\ref{eq:s}). 
Hence, $S(t)$ in (\ref{eq:s})
is independent, at all time $t$, of the dimension $d$ as well as the 
speed distribution $W(v)$ -- which we have also checked numerically (see \cite{supmat}). 

The universal result (\ref{eq:s}) is derived assuming $p(\tau)$ is exponential.
Does this result hold for other flight time PDF's $p(\tau)$? 
For non-exponential $p(\tau)$ it is difficult to compute 
$S(t)$ exactly for all $t$. With our method, this amounts to compute the 
survival probability of an effective $1d$ RW of $n$ steps where 
the last jump (corresponding to the last incomplete run in the original RTP)
differs from the $(n-1)$ first ones (the complete runs of the RTP). 
For the exponential jump distribution with rate $\gamma$, the weight of the last jump differs from
the $(n-1)$ first ones by a constant pre-factor $\gamma$ [see Eq. (\ref{joint_tau_n.1})] 
and we can still use the SA theorem, which requires an identical jump distribution for each step.
Unfortunately, for other $p(\tau)$, this trick can not be used and the SA theorem can no longer be applied. Our numerical simulations in the inset of Fig. \ref{fig:s} indeed 
indicate that $S(t)$ is no longer given by (\ref{eq:s}) for non-exponential $p(\tau)$. For such distributions, even if computing the exact expression of $S(t)$ for any finite $t$ seems challenging, it is reasonable to expect that the RTP and the aforementioned ``wait-then-jump'' model~\cite{artuso14} behave, at late times, in a qualitatively similar way. In particular, the survival probability should decay, at large time $t$, as $S(t) \propto t^{-\theta}$ with the same exponent $\theta$ for both models. From the ``wait-then-jump'' model, one can then show~\cite{artuso14} (see also \cite{supmat}) that $\theta = 1/2$ if $p(\tau)$ admits a well defined first-moment while, if the first moment is not defined, e.g. for $p(\tau) \propto \tau^{-1-\mu}$ for large $\tau$ with $\mu<1$, then $\theta=\mu/2$. In the inset of Fig. \ref{fig:s} we numerically verify these predictions for $\theta$ for the RTP with different $p(\tau)$, finding a good agreement.

As an interesting application, our universal result for $S(t)$ with an exponential $p(\tau)$ in Eq. (\ref{eq:s})
can further be used to derive the universal properties of other interesting observables for the $x$-component process of the $d$-dimensional RTP. For instance, we show in \cite{supmat} that the statistics of the number of lower records $S_N(t)$ in time $t$ for this effective $1$-d process is also universal for all $t$ and can be computed exactly. The statistics of the number of records is an important problem with a variety of applications ranging from climate science to finance \cite{Record_review}, but with very few exact analytical results. Here we show that the record statistics 
in the RTP problem is not only exactly solvable but is also universal. For example, we show that the mean number of lower records $\langle N(t) \rangle$ at all times $t$ is given by the universal formula \cite{supmat} 
\begin{eqnarray}\label{av_records}
\langle N(t) \rangle= \frac{e^{-\frac{\gamma t}{2}}}{2}  \left((2 \gamma t+3) I_0\left(\frac{t}{2}\right)+(2 \gamma t+1)
   I_1\left(\frac{\gamma t}{2}\right)\right) \;. \nonumber
\end{eqnarray}


To conclude, we computed exactly the probability $S(t)$ that the $x$-component of an RTP in $d$-dimensions does not cross the origin up to 
time $t$. For an RTP with a constant tumbling rate, we demonstrated that $S(t)$ is remarkably universal at all $t$, i.e., independent of $d$ as well as the speed distribution $W(v)$. These results are used to further compute the universal record statistics for an RTP in $d$-dimensions. It would be interesting to see if such universality extends to other other observables in RTP as well as to other models of active self-propelled particles.

\vspace*{0.3cm}

\acknowledgments{We would like to thank A. Dhar and A. Kundu for stimulating discussions at the earliest stage of this work.}



\begin{thebibliography}{26}
\bibitem{footnote_andrea}
It is interesting to point out that the joint probability distribution of the position and velocity $P(\vec x, \vec v,t)$ at time $t$
of this generalized RTP can be interpreted as a Wigner function for the problem of a quantum particle under continuous monitoring \cite{CTD19}.
\bibitem{footnote_BM}
Note that for BM, $S(t)$ is non-zero only if the particle starts initially away from the origin.
\end{thebibliography}
\end{document}


\title{Universal survival probability for a $d$-dimensional run-and-tumble particle: supplemental material}
\author{Francesco Mori }
\affiliation{LPTMS, CNRS, Univ. Paris-Sud, Universit\'e Paris-Saclay, 91405 Orsay, France}
  \author{Pierre Le Doussal}
  \affiliation{Laboratoire de Physique de l'Ecole Normale Sup\'erieure, PSL University, CNRS, Sorbonne Universit\'es, 24 rue Lhomond, 75231 Paris, France}
\author{Satya N. Majumdar}
\affiliation{LPTMS, CNRS, Univ. Paris-Sud, Universit\'e Paris-Saclay, 91405 Orsay, France}
\author{Gr\'egory Schehr }
\affiliation{LPTMS, CNRS, Univ. Paris-Sud, Universit\'e Paris-Saclay, 91405 Orsay, France}

\date{\today}

\begin{abstract}
We give the principal details of the calculations described in the main text of the Letter.
\end{abstract}

\pacs{}

\maketitle 

\section{Derivation of the survival probability $S(t)$} \label{sec:S_t}
In this section we give the details of the derivation of Eq. ($1$) in the main text, i.e. of the probability $S(t)$ that the $x$-component of the run-and-tumble particle (RTP) does not take negative values up to time $t$. We perform the computation in the most general setup, where the velocity $v$ is drawn after each tumbling from a distribution $W(v)$ with positive support and normalized to unity. It is possible to recover the usual RTP model by setting $W(v)=\delta(v-v_0)$. We consider a single RTP moving in $d$-dimensions, starting at the origin $O$ and evolving for a total time $t$.
The particle initially chooses a random direction and a random velocity $v_1$ and moves ballistically in that direction during a random time interval $\tau_1$ that is drawn from an exponential distribution 
$p(\tau)=\gamma\, e^{-\gamma \tau}$. The distance travelled during this flight $l_1=v_1 \tau_1$ is thus also a random variable. After that, the particle tumbles
instantaneously, i.e., it chooses
randomly a new direction and a new velocity. Then, it moves ballistically in that direction for an exponentially distributed time $\tau_2$
drawn independently from the same distribution $p(\tau)= \gamma\, e^{-\gamma \tau}$ and so on. More precisely, in a small time interval $dt$:

\begin{itemize}

\item With probability $\gamma\, dt$, the particle changes its direction of motion and velocity
randomly. 

\item With the complementary probability $(1-\gamma dt)$, the particle retains its direction
and moves forward in that direction by a distance $v\, dt$, where $v$ is the constant velocity of the current flight.

\end{itemize}
Note that the number $n$ of tumblings is also random. We consider the starting point $O$ as a tumbling. Thus, we always have $n\ge 1$. As explained in the letter, the last time interval $\tau_n$ will not be completed yet. Consequently, its distribution
is given by the probability $e^{-\gamma\, \tau_n}$ that no tumbling happens during the interval $\tau_n$. At variance with the previous intervals, each of which is distributed independently  according to the normalized distribution $p(\tau)= \gamma\, e^{-\gamma\, \tau}$, the distribution of $\tau_n$ is not normalized to unity.
Hence, the joint distribution of the time intervals $\{\tau_i\}=\{\tau_1,\tau_2,\ldots, \tau_n\}$ {\em and}
the number of tumblings $n$, for a fixed duration $t$ of the particle, is given by
\begin{equation}
P\left(\{\tau_i\},\, n|t\right)= 
\left[\prod_{i=1}^{n-1} \gamma\, e^{-\gamma\, \tau_i}\right]\, 
e^{-\gamma\, \tau_n}\, \delta\left( \sum_{i=1}^n \tau_i-t\right)\,.
\label{joint_tau_n.1}
\end{equation}
Let $\{l_i\}=\{l_1,l_2,\ldots, l_n\}$ denote the
straight distances travelled by the particle up to time $t$ and $\{v_i\}=\{v_1,v_2,\ldots,v_n\}$ the magnitude of the velocities in each flight.
Clearly $l_i= v_i\, \tau_i$ and $l_i\ge 0$ for all $i$. Thus, using Eq. (\ref{joint_tau_n.1}), 
the joint distribution of $\{l_i\}$, $\{v_i\}$ and the number of tumblings $n$ is given by
\begin{equation}
P\left(\{l_i\},\,\{v_i\},\, n|t\right)= \frac{1}{\gamma}\,
\left[\prod_{i=1}^{n} W\left(v_i\right) \frac{\gamma}{v_i}\, e^{-\gamma\, l_i/v_i}\right]\,
\delta\left( t- \sum_{i=1}^n \frac{l_i}{v_i}\right)\,.
\label{joint_l_v_n.1}
\end{equation}
By integrating over the speed variables $\{v_i\}$ we obtain the joint distribution of $\{l_i\}$ and $n$:
\begin{equation}
P\left(\{l_i\},\, n|t\right)= \frac{1}{\gamma}\,\int_{0}^{\infty}dv_1\,\ldots\int_{0}^{\infty}dv_n\,
\left[\prod_{i=1}^{n} \, W\left(v_i\right) \frac{\gamma}{v_i}\, e^{-\gamma\, l_i/v_i}\right]\,
\delta\left( t- \sum_{i=1}^n \frac{l_i}{v_i}\right).
\label{joint_l_n.1}
\end{equation}
As explained in the main text, the joint distribution of the $x$-components of the random vectors $\{{\vec l_i}\}$
with given norms $\{l_i\}$ can be written as (see Sec. \ref{sec:f_d} for the derivation of this result):
\begin{equation}
P\left(\{x_i\}|\{l_i\}\right)= \prod_{i=1}^n P_d(x_i | l_i) =  \prod_{i=1}^n \frac{1}{l_i} 
f_d\left(\frac{x_i}{l_i}\right)\, 
\label{x_comp_cond.1}
\end{equation}
where \begin{equation} 
f_d(z)= \frac{\Gamma(d/2)}{\sqrt{\pi}\, \Gamma((d-1)/2)}\, (1-z^2)^{(d-3)/2}\, \theta(1-|z|)\, .
\label{fdz.1}
\end{equation}
Here $\Gamma(y)$ is the Gamma function and $\theta(y)$ is the Heaviside step function: $\theta(y)=1$ if $y\ge 0$ and $\theta(y)=0$ if $y<0$.
We can then write down explicitly the joint distribution of the $x$-components $\{x_i\}$, the norms $\{l_i\}$ and the number of tumblings $n$ at fixed total time $t$ as 
\begin{eqnarray}
P\left(\{x_i\}, \{l_i\}, n|t\right)& = & 
P\left(\{x_i\}|\{l_i\}\right)\, P\left(\{l_i\},\, n|t\right)
\nonumber \\
&=& \frac{1}{\gamma}\int_{0}^{\infty}dv_1\,\ldots\int_{0}^{\infty}dv_n\, \left[ \prod_{i=1}^n \frac{1}{l_i}
f_d\left(\frac{x_i}{l_i}\right)\,W(v_i) \frac{\gamma}{v_i}\, e^{-\gamma\, l_i/v_i}\right]\,
\delta\left( t- \sum_{i=1}^n \frac{l_i}{v_i}\right),
\label{full_joint.1}
\end{eqnarray}
where we used the results in Eqs. (\ref{joint_l_n.1}) and (\ref{x_comp_cond.1}).
Having obtained this joint distribution we can now integrate over the $\{l_i\}$ variables to
obtain the marginal joint distributions of $\{x_i\}$ and $n$, given $t$
\begin{equation}
P\left(\{x_i\},\, n|t\right)= \frac{1}{\gamma}\, \int_{0}^{\infty} dl_1\ldots \int_0^{\infty}
dl_n\, \int_{0}^{\infty}dv_1\,\ldots\int_{0}^{\infty}dv_n\, \left[ \prod_{i=1}^n \frac{1}{l_i}
f_d\left(\frac{x_i}{l_i}\right)\, W\left(v_i\right)\frac{\gamma}{v_i}\, e^{-\gamma\, l_i/v_i}\right]\,
\delta\left( t- \sum_{i=1}^n \frac{l_i}{v_i}\right)\, .
\label{joint_xn.1}
\end{equation}
The result in Eq. (\ref{joint_xn.1})  then provides us an effective $x$-component process $\{x_i\}$ projected from the $d$-dimensional RTP
of fixed duration $t$. To further simplify this $x$-component process, we take a Laplace transform
with respect to $t$ that decouples the integrals over the $\{l_i\}$ variables
\begin{eqnarray}
\int_0^{\infty} dt\, e^{-s\,t}\, P\left(\{x_i\},\, n|t\right) &=&
\frac{1}{\gamma}\, \int_{0}^{\infty} dl_1\ldots \int_0^{\infty}
dl_n\, \int_{0}^{\infty}dv_1\,\ldots\int_{0}^{\infty}dv_n\,\left[ \prod_{i=1}^n \frac{1}{l_i}
f_d\left(\frac{x_i}{l_i}\right)\, W\left(v_i\right) \frac{\gamma}{v_i}\, e^{-(\gamma+s)\, l_i/v_i}\right] \nonumber \\
&=& \frac{1}{\gamma}\, \left(\frac{\gamma}{\gamma+s}\right)^n \, 
\prod_{i=1}^n \int_0^{\infty} 
\frac{dl_i}{l_i} f_d\left(\frac{x_i}{l_i}\right)\,\int_{0}^{\infty}dv_i\,W\left(v_i\right) 
\frac{(\gamma+s)}{v_i}\, e^{-(\gamma+s)\,l_i/v_i} \nonumber \\
&=& \frac{1}{\gamma}\, \left(\frac{\gamma}{\gamma+s}\right)^n \prod_{i=1}^n {\tilde p}_s(x_i)
\label{lt_joint.1}
\end{eqnarray}
where we have defined
\begin{equation}
{\tilde p}_s(x)= \int_0^{\infty} \frac{dl}{l} f_d\left(\frac{x}{l}\right)\, \int_{0}^{\infty}dv\,  W\left(v\right)
\frac{(\gamma+s)}{v}\, e^{-(\gamma+s)\,l/v}\, .
\label{ptilde.1}
\end{equation}
Note that in getting from the first to the second line above, we 
have multiplied and divided by a factor $(\gamma+s)^n$ so that the function ${\tilde p}_s(x)$, which depends on the parameters $s$, $d$, $\gamma$ and on the speed distribution $W(v)$, can be interpreted as a probability density 
function (PDF) of a random variable $x$.
Manifestly ${\tilde p}_s(x)$ is non-negative and normalized to unity. Indeed, integrating over $x$ one gets
\begin{eqnarray}
\int_{-\infty}^{\infty} {\tilde p}_s(x)\, dx &= & (\gamma+s)\,\int_0^{\infty} dl\, \int_{-\infty}^{\infty} \frac{dx}{l}\, f_d\left(\frac{x}{l}\right)\, \int_{0}^{\infty}\frac{dv}{v}W\left(v\right)
e^{-(\gamma+s)\, l/v}\, \nonumber \\
&=&  \left(\gamma+s\right)\,\int_0^{\infty} \frac{dv}{v}W\left(v\right)\int_0^{\infty} dl\, e^{-(\gamma+s)\, l/v} \int_{-1}^1 dz\, f_d(z) \nonumber \\
&=&\int_0^{\infty} dv\,W\left(v\right) \,=1 ,
\label{normal_psx}
\end{eqnarray}
where we used the fact that $f_d(z)$ given in Eq. (\ref{fdz.1})
is supported over the finite interval $z\in [-1,1]$ and is normalized to unity and that $W(v)$ is normalized to unity. As we will see below, the precise expression for ${\tilde p}_s(x)$ is not relevant, as long as it is continuous and symmetric in $x$. {Note that this property for ${\tilde p}_s(x)$ will hold for general factorized jump distributions $P\left(\{x_i\}|\{l_i\}\right) = \prod_{i=1}^n P_d(x_i|l_i)$ as in Eq. (\ref{x_comp_cond.1}), provided the conditional distribution $P_d(x | l)$ is symmetric in $x$, i.e. $P_d(x | l) = P_d(-x | l)$.} Finally, inverting the Laplace transform in Eq. (\ref{lt_joint.1}) formally, we have the joint distribution of $\{x_i\}$ and $n$ for a fixed $t$
\begin{equation}
P\left(\{x_i\}, n|t\right)=  
\int \frac{ds}{2\pi\,i} e^{s\, t}\, 
\frac{1}{\gamma}\, \left(\frac{\gamma}{\gamma+s}\right)^n \prod_{i=1}^n {\tilde p}_s(x_i)\, ,
\label{Pxn.1}
\end{equation}
where the integral is over the Bromwich contour (imaginary axis in this case) in the complex $s$ plane. 

\begin{figure}
\includegraphics[width = 0.5\linewidth]{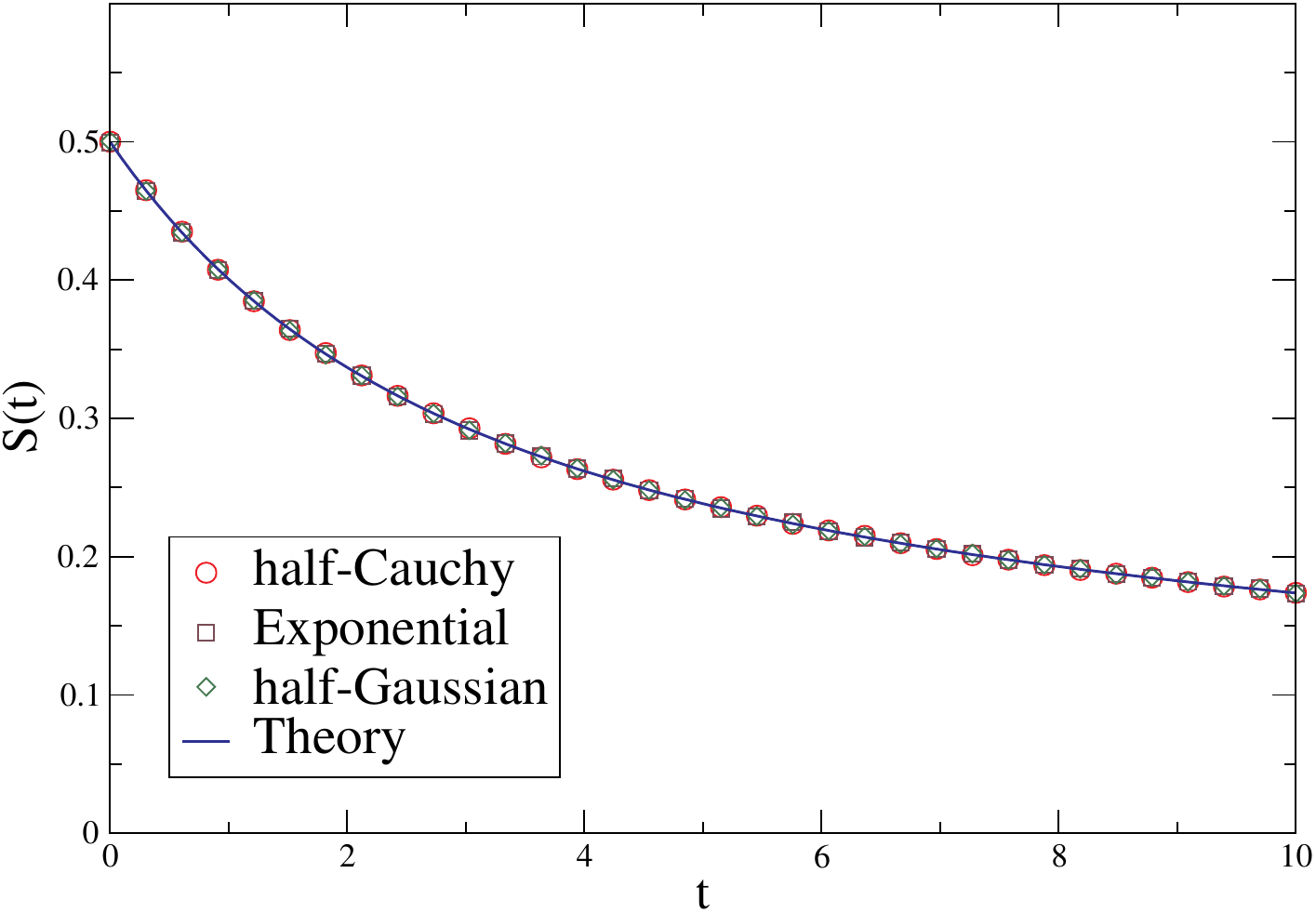}
\caption{Plot of $S(t)$ evaluated numerically  
for different velocity distributions $W(v)$ and an exponential distribution $p(\tau)$ in dimension $d=2$. The solid line corresponds to the
exact analytical result in Eq. (\ref{surv.2}).} \label{Fig_SM_Wv}
\end{figure}

The survival probability $S(t)$ of this $x$-component process up to time $t$ is the probability of the event that the successive sums $x_1$, $x_1+x_2$, $\ldots$, $x_1+x_2+\ldots x_n$ are all positive. We recall that the number of tumblings $n$ is also a random variable. Thus, summing over $n$ one obtains
\begin{equation}
S(t) = \sum_{n=1}^{\infty} \int_{-\infty}^{\infty} dx_1\ldots \int_{-\infty}^{\infty} dx_n\,
\left[\theta(x_1)\,\theta(x_1+x_2)
\ldots  \theta(x_1+x_2+\ldots+x_n)\right]\, P\left(\{x_i\}, n|t\right)\, ,
\label{St.1}
\end{equation}
where $ P\left(\{x_i\}, n|t\right)$ is the joint distribution of $\{x_i\}$ and $n$ for 
fixed $t$. Plugging the expression for $P\left(\{x_i\}, n|t\right)$ given in Eq. (\ref{Pxn.1})
gives
\begin{eqnarray}
S(t) & = & \int \frac{ds}{2\pi\,i} e^{s\, t}\,
\frac{1}{\gamma}\, \sum_{n=1}^{\infty} \left(\frac{\gamma}{\gamma+s}\right)^n\,
\int_{-\infty}^{\infty} dx_1\ldots \int_{-\infty}^{\infty} dx_n\,
\left[\theta(x_1)\,\theta(x_1+x_2)
\ldots  \theta(x_1+x_2+\ldots+x_n)\right]\, \prod_{i=1}^n {\tilde p}_s(x_i) \nonumber \\
& = & \int \frac{ds}{2\pi\,i} e^{s\, t}\,
\frac{1}{\gamma}\, \sum_{n=1}^{\infty} \left(\frac{\gamma}{\gamma+s}\right)^n\, q_n\,,
\label{St.2}
\end{eqnarray}
where we have defined the multiple integral
\begin{equation}
q_n= \int_{-\infty}^{\infty} dx_1\ldots \int_{-\infty}^{\infty} dx_n\,
\left[\theta(x_1)\,\theta(x_1+x_2)
\ldots  \theta(x_1+x_2+\ldots+x_n)\right]\, \prod_{i=1}^n {\tilde p}_s(x_i)\, .
\label{qn.1}
\end{equation}
However, as mentioned in the text, this quantity $q_n$ in Eq. (\ref{qn.1}) has the following interpretation.
Consider a one-dimensional discrete-time random walk, starting at the origin $x=0$ and making independent jumps
at each step with jump length drawn from the PDF
${\tilde p_s}(x)$. Then $q_n$ is just the probability that the walker does not visit the negative axis up to step $n$. 
Notably, since ${\tilde p}_s(x)$ is continuous and symmetric, the Sparre Andersen theorem \cite{SA_54} states 
that $q_n$ is universal, i.e., independent of ${\tilde p}_s(x)$ and is given by:
\begin{equation}
q_n= {2n \choose n}\, 2^{-2n}\, \quad\quad\quad n=0,1,2,\ldots
\label{sa.1}
\end{equation}
Note that this formula is valid for any $n$, and hence the universality holds for all $n$, and not 
just asymptotically for large $n$. The generating function of $q_n$ is thus also universal
\begin{equation}
\sum_{n=0}^{\infty} q_n\, z^n = \sum_{n=0}^{\infty}  {2n \choose n}\, 
\left(\frac{z}{4}\right)^n= \frac{1}{\sqrt{1-z}}\, .
\label{sa.2}
\end{equation}
Using this result (\ref{sa.2}) in Eq. (\ref{St.2}) and noticing that the sum in Eq. (\ref{St.2}) does not include the $n=0$ term leads to the result  
\begin{equation}
S(t)= \int \frac{ds}{2\pi\,i} e^{s\, t}\,
\frac{1}{\gamma}\ \left[\sqrt{\frac{\gamma+s}{s}}-1\right]\, ,
\label{sa.3}
\end{equation}
which is Eq. (13) in the main text. 
Note that this result is universal in the sense that it does not depend on the dimension $d$ or on the speed 
distribution $W(v)$. Indeed, $W(v)$ and $d$ appear only in Eq. (\ref{St.2}) through the PDF ${\tilde p}_s(x)$. 
However, we have seen that as a consequence of the Sparre Andersen theorem the result is completely independent of the 
specific form of ${\tilde p}_s(x)$, as long as it is symmetric and continuous. 
As explained in the main text, the Laplace inversion in Eq. (\ref{sa.3}) 
can be explicitly performed leading to 
\begin{equation}
S(t)= \frac{1}{2}\, e^{-\gamma t/2}\, \left[I_0(\gamma t/2)+ I_1(\gamma t/2)\right]\,
\label{surv.2}
\end{equation}
where $I_0(z)$ and $I_1(z)$ are modified Bessel functions, as given in Eq. (1) in the main text. In Fig. \ref{Fig_SM_Wv} we show 
a plot of $S(t)$ evaluated numerically  
for different velocity distributions $W(v)$ and an exponential distribution $p(\tau)$ in dimension $d=2$, which shows a very good agreement with our exact result (\ref{surv.2}).

\section{Derivation of the formula in Eq. (\ref{fdz.1}) for the marginal distribution $P_d(x|l)$} \label{sec:f_d}

We consider a random vector $\vec l$ of fixed magnitude $l$ in $d$-dimensions and compute
the marginal distribution $P_d(x|L)$ of its $x$-component, given fixed $l$. The PDF of a random vector $\vec l$
of fixed magnitude $l$ is simply
\begin{equation}
P(\vec l)= \frac{1}{S_d\, l^{d-1}}\, \delta\left(|\vec l|-l\right)\,  ,
\label{pdf_vecl.A1}
\end{equation}
where 
\begin{equation}
S_d= \frac{2 \pi^{d/2}}{\Gamma(d/2)}\, 
\end{equation}
is just the surface area of a $d$-dimensional sphere of unit radius.
It is convenient to rewrite Eq. (\ref{pdf_vecl.A1}) as
\begin{equation}
P(\vec l)= \frac{2}{S_d\, l^{d-2}}\, \delta\left(|\vec l|^2- l^2\right)\, .
\label{pdf_vecl.A2}
\end{equation}
Let $|\vec l|^2= z_1^2+z_2^2+\ldots z_d^2$ where $z_k$ denotes the component of the vector 
$\vec l$ along the $k$-th direction. Therefore, the marginal distribution $P_d(x|l)$, 
for instance along the $x$ direction,
is obtained by keeping $z_1=x$ fixed while integrating over the other components
\begin{eqnarray}
P_d(x|l) & =&  \int P(\vec l)\, \delta(z_1-x)\, dz_1\, dz_2\, \ldots dz_d \nonumber \\
& =& \frac{2}{S_d\, l^{d-2}}\, 
\int \delta\left(z_2^2+z_3^2+\ldots +z_d^2- (l^2-x^2)\right)\, dz_2\, dz_3\, \ldots dz_d\, ,
\label{marg_dist.1}
\end{eqnarray}
where we used Eq. (\ref{pdf_vecl.A2}) in going from the first to the second line above.
Let $R^2= z_2^2+z_3^2+\ldots+z_d^2$. Then the $(d-1)$-dimensional integral in Eq. (\ref{marg_dist.1})
can be performed in the radial coordinate
\begin{equation}
P_d(x|L)  = \frac{2\, S_{d-1}}{S_d\, l^{d-2}}\, \int_0^{\infty} \delta\left(R^2- (l^2-x^2)\right)\, R^{d-2}\, dR  
\label{marg_dist.2}
\end{equation}
where we recall $S_{d-1}$ is the surface area of a $(d-1)$-dimensional unit sphere. The single
radial integral in Eq. (\ref{marg_dist.2}) can be trivially done by making a change of variable $R^2=u$
\begin{eqnarray}
P_d(x|l) &=& \frac{S_{d-1}}{S_d\, l^{d-2}}\, 
\int_0^{\infty} \delta\left(u-(l^2-x^2)\right)\, u^{(d-3)/2}\, du \nonumber \\
& = &  \frac{S_{d-1}}{S_d\, l^{d-2}}\, (l^2-x^2)^{(d-3)/2}\, \theta(l-|x|)\, .
\label{marg_dist.3}
\end{eqnarray}
Using the formula for $S_d$ in Eq. (\ref{pdf_vecl.A1}) and rearranging the terms, we get
\begin{equation}
P_d(x|l)= \frac{1}{l}\, f_d\left(\frac{x}{l}\right)\, ,
\label{fdz.A1}
\end{equation}
where
\begin{equation} \label{fdz.A2}
f_d(z)= \frac{\Gamma(d/2)}{\sqrt{\pi}\, \Gamma((d-1)/2)}\, (1-z^2)^{(d-3)/2}\, \theta(1-|z|)\,,
\end{equation}
as given in Eq. (4) in the main text.
One can check easily that $f_d(z)$ is normalized to unity over the support $z\in [-1,1]$.

\section{Survival probability for random walks with spatio-temporal correlations}
As mentioned in the main text, the RTP model is a particular case of a random walk (RW) with spatio-temporal correlation. Let us consider a RW on the real line. Let
\begin{equation}
(x_{1},\tau_{1}),\,(x_{2},\tau_{2}),\ldots,(x_{j},\tau_{j}),\ldots
\end{equation}
be a sequence of independent identically distributed pairs of random
variables corresponding to the step length $x_{i}$ and the associated
time $\tau_{i}$. We assume that each pair $(x_{i},\tau_{i})$ is distributed
according to the some joint PDF $p(x,\tau),$
which is assumed to be continuous in $x$ and spatially symmetric:
$p(x,\tau)=p(-x,\tau).$ 
After $n$ steps the RW will be in position $X_{n}$ at
time $T_{n}$, where
\begin{equation}
X_{n}=\sum_{k=0}^{n}\,x_{k},\quad\quad T_{n}=\sum_{k=0}^{n}\,\tau_{k}\,.\label{eq:S_n}
\end{equation}

To study the probability $S(t)$ that the negative side of the $x$-axis is not reached up to time $t$ we 
also need to specify how the walker moves when taking a step. One possibility is that in order to take a step $x_i$ in a time $\tau_i$ the walker moves with constant velocity $v_i=x_i/\tau_i$, as in the case of the RTP. 
Another possibility is that the walker remains in its position for a time 
$\tau_i$ and then takes an instantaneous jump $x_i$. 
It turns out that for this latter ``wait-then-jump model''
the survival probability $S(t)$ can be computed exactly for any distribution $p(x,\tau)$. This result was recently 
obtained by Artuso et al. using a combinatorial lemma \cite{artuso14} combined
with the Sparre Andersen theorem. 
Here we propose an alternative non-combinatorial derivation based on our 
technique presented in Sec. \ref{sec:S_t}.
It is convenient to derive the probability $\pi(t)$ that the first entrance to the 
negative axis happens at time $t$. One can then obtain the survival probability $S(t)$ using the relation 
\begin{equation} \label{eq:relation_survival}
S'(t)=-\pi(t)\,.
\end{equation}
In fact, Artuso et. al.~\cite{artuso14} computed exactly the Laplace transform of $\pi(t)$, rather than of $S(t)$, but
these two are simply related due to the relation (\ref{eq:relation_survival}). In order to compare to the
result of the RTP in Eq. (\ref{surv.2}), we will compute $S(t)$ for the ``wait-then-jump model''.  

Consider a trajectory of the 
``wait-then-jump model'' up to time $t$. By definition, in the
``wait-then-jump model'' there are $n$ complete jumps such that $\sum_{i=1}^n \tau_i=t$, where
each pair $(x_i,\tau_i)$ is drawn independently from the PDF $p(x,\tau)$. This corresponds to imposing that
there is a jump at time $t$. Note that the number of jumps $n$ in time $t$
is a random variable, as in the RTP.
The joint PDF of $\{(x_{i},\tau_{i})\}_{1\leq i\leq n}$ and of $n$,
at fixed total time $t$ is then given by
\begin{equation}
P(\{(x_{i},\tau_{i})\},n|t)=\prod_{i=1}^{n}p(x_{i},\tau_{i})\,\delta\left(\sum_{i=1}^{n}\tau_{i}-t\right),
\end{equation}
where the delta function enforces the constraint on the total time.
We integrate over the $\tau$ variables to obtain the marginal of $\{x_{i}\}$
and $n$
\begin{equation}
P(\{x_{i}\},n|t)=\int_{0}^{\infty}d\tau_{1}\,\ldots\int_{0}^{\infty}d\tau_{n\,}\prod_{i=1}^{n}p(x_{i},\tau_{i})\,\delta\left(\sum_{i=1}^{n}\tau_{i}-t\right)\,.
\end{equation}
Taking a Laplace transform with respect to $t$ we decouple the integrals
over the $\tau$ variables
\begin{equation}
\int_{0}^{\infty}dt\,P(\{x_{i}\},n|t)\,e^{-st}=\prod_{i=1}^{n}\int_{0}^{\infty}d\tau_{i}p(x_{i},\tau_{i})\,e^{-s\,\tau_{i}}\,.\label{eq:marginal_Laplace}
\end{equation}
It is useful to rewrite the right-hand side of Eq. (\ref{eq:marginal_Laplace})
as 
\begin{equation}
\int_{0}^{\infty}dt\,P(\{x_{i}\},n|t)\,e^{-st}=c(s)^{n}\prod_{i=1}^{n}\tilde{p}_{s}(x_{i})\,,\label{eq:marginal_Laplace_2}
\end{equation}
where $c(s)$ is defined as
\begin{equation}\label{eq:c_s}
c(s)=\int_0^{\infty}d\tau\,\int_{-\infty}^{\infty}dx\,p(x,\tau)\,e^{-s\tau}
\end{equation}
 and 
\begin{equation}
\tilde{p}_{s}(x)=\frac{1}{c(s)}\int_{0}^{\infty}d\tau\,p(x,\tau)\,e^{-s\tau}\,.
\end{equation}
Note that $\tilde{p}_{s}(x)$ can be interpreted as a probability
density function. Indeed, it is clearly non-negative and normalized to unity.
Moreover, since we assume $p(x,\tau)$ to be continuous and symmetric
with respect to $x$, $\tilde{p}_{s}(x)$ will also be continuous
and symmetric. The probability that the walker enters in the positive
axis for the first time at time $t$ can be simply written as
\begin{equation}
\pi(t)=\sum_{n=1}^{\infty}\int_{0}^{\infty}dx_{1}\ldots\int_{0}^{\infty}dx_{n}\,P(\{x_{i}\},n|t)\:\Theta_{n}(\{x_{i}\})\,,\label{eq:pi_t-1}
\end{equation}
with the notation 
\begin{equation}
\Theta_{n}(\{x_{i}\})=\theta(X_{1})\,\theta(X_{2})\,\ldots\theta(X_{n-1})\theta(-X_{n}),
\end{equation}
where $X_{i}$'s are the partial sums defined in Eq. (\ref{eq:S_n}) and
$\theta(x)$ is the Heaviside step function: $\theta(x)=1$ if $x>0$
and $\theta(x)=0$ otherwise. In other words, $\Theta_{n}(\{x_{i}\})$
enforces that the negative axis is reached for the first time at step
$n$. Taking a Laplace transform of both sides of Eq. (\ref{eq:pi_t-1})
and using Eq. (\ref{eq:marginal_Laplace_2}), we obtain
\begin{equation}
\hat{\pi}(s)=\sum_{n=1}^{\infty}c(s)^{n}f_{n}\,,\label{eq:pi_s}
\end{equation}
where
\begin{equation}
f_{n}=\int_{0}^{\infty}dx_{1}\ldots\int_{0}^{\infty}dx_{n}\prod_{i=1}^{n}\tilde{p}_{s}(x_{i})\Theta_{n}(\{x_{i}\})\,.
\end{equation}
Notably, $f_{n}$ can be interpreted at the probability of first passage to the negative $x$-axis for a RW 
\begin{equation}
X_{k}=X_{k-1}+\eta_{k}\,\quad\quad y_{0}=0\,,
\end{equation}
where $\eta_{k}$ is a random number extracted from the PDF $\tilde{p}_{s}(x_{i})$. Then, since $\tilde{p}_{s}(x_{i})$
is continuous and symmetric, according to the Sparre Andersen theorem $f_{n}$ is universal and its generating function 
can be computed as follows. Clearly, $f_n= q_{n-1}-q_n$ for $n\ge 1$, where $q_n$ is the probability that the
random walker stays positive up to step $n$. Taking a generating function, we get
\begin{equation}
\sum_{n=1}^{\infty} f_n\, z^n= \sum_{n=1}^{\infty} \left[q_{n-1}- q_n\right]\, z^n= 1- (1-z) \sum_{n=0}^{\infty} q_n\, z^n\, ,
\label{fnqn.1}
\end{equation}
where we used $q_0=1$. Since, ${\tilde p}_s(x)$ is a continuous and symmetric PDF, the Sparre Andersen theorem
can be applied which states that $\sum_{n=0}^{\infty} q_n\, z^n= 1/\sqrt{1-z}$. Hence, from Eq. (\ref{fnqn.1}) one gets
\begin{equation}
\sum_{n=1}^{\infty}f_{n}\, z^n=1-\sqrt{1-z}\, .
\label{eq:sa_theorem}
\end{equation}
Using this result (\ref{eq:sa_theorem}) in Eq. (\ref{eq:pi_s}) we obtain
\begin{equation}\label{eq:final_pi}
\hat{\pi}(s)=1-\sqrt{1-c(s)}\,.
\end{equation}
where $c(s)$ is given in Eq. (\ref{eq:c_s}). This is indeed the result of Artuso et. al.~\cite{artuso14}
obtained originally using combinatorial method. Our derivation above is non-combinatorial and a bit simpler
in our opinion. 

From Eq. (\ref{eq:final_pi}) one can compute the Laplace 
transform of the survival probability $S(t)$. Indeed, using Eq. (\ref{eq:relation_survival}) it is easy to show that
\begin{equation}
\tilde{S}(s)=\int_0^\infty\,dt\,S(t)\,e^{-st}=\frac{1-\tilde{\pi}(s)}{s}\,.
\end{equation}
Using Eq. (\ref{eq:final_pi}) we obtain that
\begin{equation}
\tilde{S}(s)=\frac{\sqrt{1-c(s)}}{s}\,,
\label{surv_final}
\end{equation}
where $c(s)$ is given in Eq. (\ref{eq:c_s}). To compare with the RTP model in $d$ dimensions, let us now choose 
\begin{equation}
p(x,\tau)=p(x|\tau)p(\tau)= \frac{1}{v_0\tau}f_d\left(\frac{x}{v_0\tau}\right)\, p(\tau)\,,
\end{equation}
where $f_d(z)$ is given in Eq. (\ref{fdz.A2}). Then  we get
from Eq. (\ref{eq:c_s}),
\begin{align}
c(s) & =\int_{0}^{\infty}d\tau\,\int_{-\infty}^{\infty}dx\:p(x,\tau)\,e^{-s\tau}\,=\int_{0}^{\infty}d\tau\,\int_{-\infty}^{\infty}dx\:\frac{1}{v_{0}\tau}f_{d}\left(\frac{x}{v_{0}\tau}\right)\,p(\tau)\,e^{-s\tau}\,\\
 & =\int_{0}^{\infty}d\tau\,e^{-s\tau}p(\tau)\,\int_{-\infty}^{\infty}dx\:\frac{1}{v_{0}\tau}f_{d}\left(\frac{x}{v_{0}\tau}\right)\,=\int_{0}^{\infty}d\tau\,e^{-s\tau}p(\tau)\,\int_{-\infty}^{\infty}dz\:f_{d}\left(z\right)\,\\
 & =\int_{0}^{\infty}d\tau\,e^{-s\tau}p(\tau)\,=\tilde{p}(s)\,.
\end{align}
where we have used the fact that $f_d(z)$ is normalized to unity in going from the second to the third line above. Note that $\tilde{p}(s)$ is simply defined as the Laplace transform of $p(\tau)$. Then, using Eq. (\ref{surv_final}), we obtain that
\begin{equation}
\tilde{S}(s)=\frac{\sqrt{1-\tilde{p}(s)}}{s}\,.
\label{surv_final_rtm}
\end{equation}
In the most relevant case of an exponential distribution $p(\tau)=\gamma\,e^{-\gamma\tau}$ one obtains that $c(s)= \gamma/(\gamma+s)$. Consequently, Eq. (\ref{surv_final_rtm}) gives
${\tilde S}(s)= 1/\sqrt{s(\gamma+s)}$. Inverting the Laplace transform explicitly, we then get the exact survival probability at all $t$ for this specific ``wait-then-jump model'' with exponential time distribution
\begin{equation}
S(t)= e^{-\gamma\, t/2}\, I_0(\gamma\, t/2)\, ,
\label{surv_final.1}
\end{equation}
where $I_0(z)$ is again the modified Bessel function. The result in (\ref{surv_final.1}) is manifestly different
from the RTP result in Eq.~(\ref{surv.2}) (also in Eq. (1) of the main text). This clearly shows that 
the exact result in Eq. (\ref{surv_final}) for the
``wait-then-jump model'' can not be used to derive our main result for the RTP in Eq. (1) of the main text. Note however that for late times the result in Eq. (\ref{surv_final.1}) has the same asymptotic behavior as the RTP result, namely $S(t)\sim 
1/\sqrt{\pi\gamma t}$.

Moreover, as explained in the main text, Eq. (\ref{surv_final_rtm}) can be useful to compute the late time behavior of $S(t)$ for the RTP model with a generic time distribution $p(\tau)$. Indeed, one expects that $S(t)\sim t^{-\theta}$ when $t\to\infty$ (and this is confirmed by our numerical simulations shown in Fig. 3 of the Letter). Moreover, for late times, it is natural to conjecture that the exponent $\theta$ is the same for the RTP model and for the ``wait-then-jump model''. Here, we compute the exponent $\theta$ for different time distributions $p(\tau)$ in the ``wait-then-jump'' setup. It is useful to distinguish two cases, depending on whether 
$p(\tau)$ has a well defined first moment or not. 
\vspace*{0.2cm}

\noindent{\it The case where $p(\tau)$ has a well defined first moment.} In this case, the Laplace transform $\tilde{p}(s)$ can be expanded, for small $s$, as
\begin{equation}\label{eq:laplace_p_finite}
\tilde p(s)\simeq 1- \langle \tau \rangle \,s+o(s)\,,
\end{equation}
where $\langle \tau \rangle=\int_{0}^\infty\,d\tau\,\tau\,p(\tau)$ is the first moment of $\tau$. Using Eq. (\ref{surv_final_rtm}) we obtain that, for small $s$
\begin{equation}
\tilde{S}(s)\sim \sqrt{\frac{\langle \tau \rangle}{s}}\,.
\end{equation}
Inverting the Laplace transform gives for late times that
\begin{equation}\label{S_av_tau}
S(t)\sim \sqrt{\frac{\langle \tau \rangle}{\pi \, t}}\,.
\end{equation}
Hence, if $\langle \tau \rangle$ is finite we obtain that $\theta=1/2$. Note that for the exponential jump distribution with rate $\gamma$, one has $\langle \tau \rangle = 1/\gamma$ and this formula (\ref{S_av_tau}) yields back $S(t) \sim 1/\sqrt{\pi \gamma t}$, as it should.

\vspace*{0.2cm}

\noindent{\it The case where $p(\tau)$ has a diverging first moment.} If the average value of $\tau$ is diverging, i.e. if $p(\tau)\sim\tau^{-\mu-1}$ for $\tau\to\infty$ with $0<\mu<1$, then $\tilde{p}(s)$ can be expanded for small $s$ as
\begin{equation}\label{eq:laplace_p_diverging}
p(s) =  1-(a\,s)^\mu+o(s^\mu)\;,
\end{equation}
where $a$ denotes a microscopic time scale. Using Eq. (\ref{surv_final_rtm}) we obtain that, when $s\to0$,
\begin{equation}
\tilde{S}(s)\sim s^{\mu/2-1}\,.
\end{equation}
Inverting the Laplace transform we get that when $t\to\infty$
\begin{equation}
S(t)\sim t^{-\mu/2}\,,
\end{equation}
and, hence, in this case $\theta=\mu/2$.

\section{Universal record statistics for the RTP}

\begin{figure}
\includegraphics[width = 0.5 \linewidth]{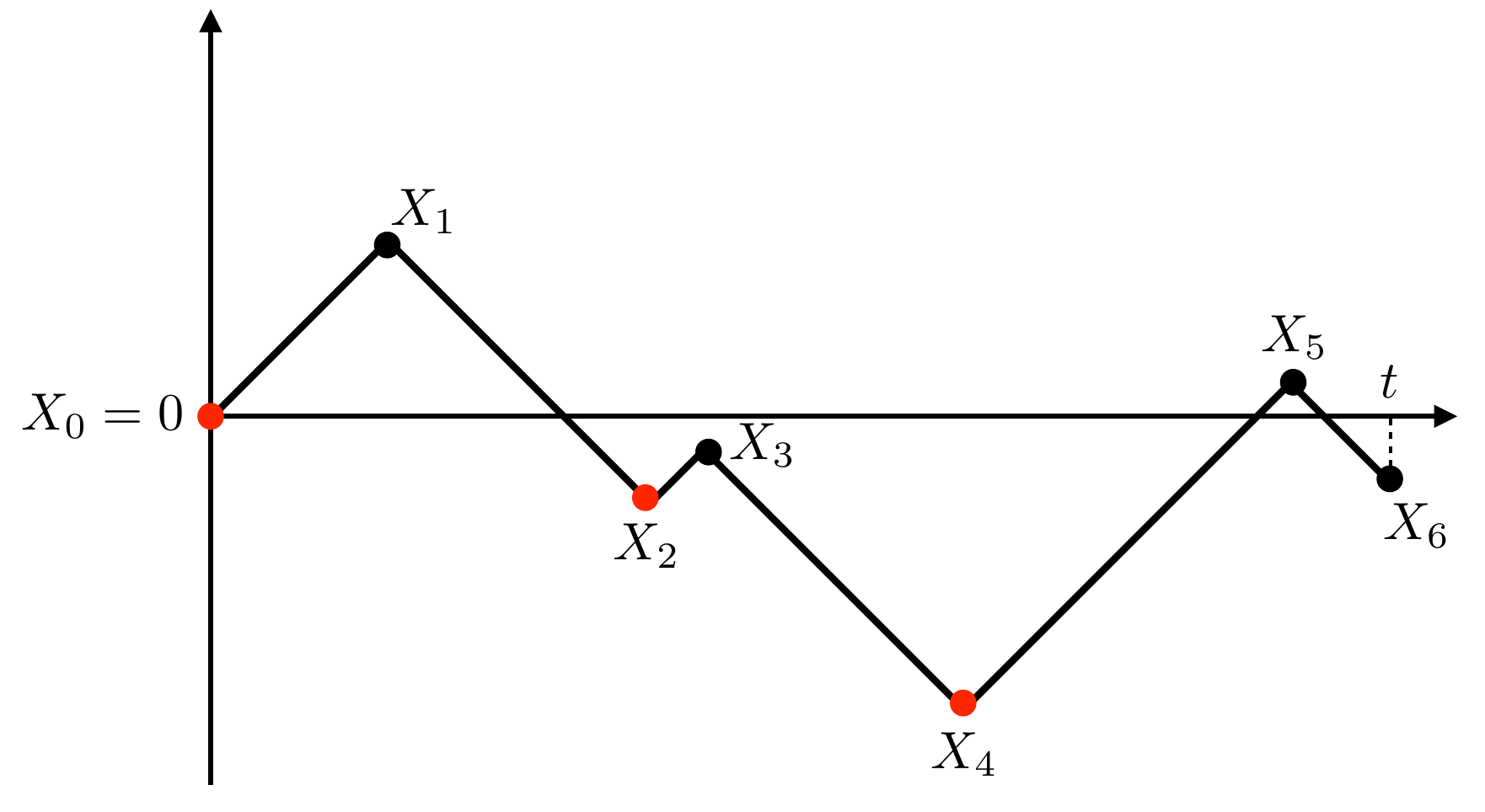}
\caption{Illustration of a trajectory of the RTP (black solid line) together with the positions of the associated random walk $X_0 = 0, X_1, \ldots X_{n}$ with $n=6$, up to time $t$ (dots). The lower records are indicated in red, the first position $X_0$ being counted as a lower record. Note that the final position $X_n$ can in principle be also a record -- although not in the above figure. The observable $S_N(t)$ is the probability that the random walk has exactly $N$ lower records up to time $t$. For $N=1$, $S_1(t)$ coincides the survival probability up to time $t$.}\label{fig_record}
\end{figure}

In this section we show that our results for the survival probability $S(t)$ for a $d$-dimensional RTP can be used to compute the statistics of records for the $x$-component of the RTP process. The universality of $S(t)$ for the RTP with an exponential distribution of the flight times (corresponding to a constant tumbling rate $\gamma$) also renders the statistics of the records for the $x$-component 
universal in this problem, i.e. independent of the dimension $d$ as well as the speed distribution $W(v)$. The statistics of records for a stochastic sequence has been extensively studied and has found many applications from climatology to finance \cite{Record_review}. In general, it is 
quite hard to obtain exact results for the record statistics for a correlated sequence. Below, we see that, using our method as detailed in the main text, we can compute the exact record statistics of the $x$-component of the RTP in $d$ dimensions and show that it is universal. This is one of the rare examples of an exact solution for the record statistics for a correlated sequence.   

Let us start by defining a record. We consider a trajectory in $d$ dimensions of the RTP of duration $t$ starting at the origin. Let $n$ denote the number of runs in this trajectory, which itself is a random variable. We now look at the $x$-components of the $n$ successive runs and denote them by $x_1, x_2, \cdots, x_n$. The $x$-component of the positions of the RTP are denoted by $X_0=0, X_1 = x_1, X_2=x_1+x_2, X_3=x_1+x_2+x_3, \ldots$ (see Fig. \ref{fig_record}). The joint distribution of the $x_i$'s and $n$ has been computed in Eq. (8) of the main text and is given by
\begin{equation}
P\left(\{x_i\}, n|t\right)=  
\int \frac{ds}{2\pi\,i} e^{s\, t}\, 
\frac{1}{\gamma}\, \left(\frac{\gamma}{\gamma+s}\right)^n \prod_{i=1}^n {\tilde p}_s(x_i) \;,
\label{Pxn.1_supp}
\end{equation} 
where $\tilde p_s(x_i)$ is given in Eq. (14) of the main text for a general speed distribution $W(v)$ and general dimension $d$. Therefore, the $X_i$'s can be viewed as the position of a one-dimensional discrete-time random walker with correlated steps given in Eq. (\ref{Pxn.1_supp}). A lower record happens at step $k$ iff the value $X_k$ is lower than all the previous values, i.e., $X_k < \min \{X_0=0, X_1, \cdots, X_{k-1} \} $ (see Fig. \ref{fig_record}). By convention, $X_0 = 0$ is a lower record. Note that the final position $X_n$ can also be a record. A natural question is then: how many records occur in time $t$? We denote by $S_N(t)$ the probability that there are exactly $N$ lower records up to time $t$. Clearly, when $N=1$ this corresponds to the event that the position has never gone below $0$ up to time $t$. But this precisely the survival probability $S(t)$ that we have computed in the main text, thus $S_1(t) = S(t)$. We can then think of $S_N(t)$ as a natural generalization of the survival probability $S(t)$. One can similarly define upper records for the $x$-component of the RTP, whose statistics are exactly identical to the lower records, due to the $x \to -x$ symmetry of the RTP. {An alternative physical picture of this record process is as follows: whenever the particle achieves a new lower record, one can imagine that the absorbing barrier gets pushed to this new record value. For example, before the second record happens the absorbing barrier is at $X_0=0$. If the second lower record happens at step $k$ with value $X_k < 0$ (for example in Fig. \ref{fig_record} the second record happens at $k=2$), the absorbing barrier gets shifted to $X_k$, till the occurence of the next lower record (see Fig. \ref{fig_record}).}

\begin{figure}
\includegraphics[width = 0.5\linewidth]{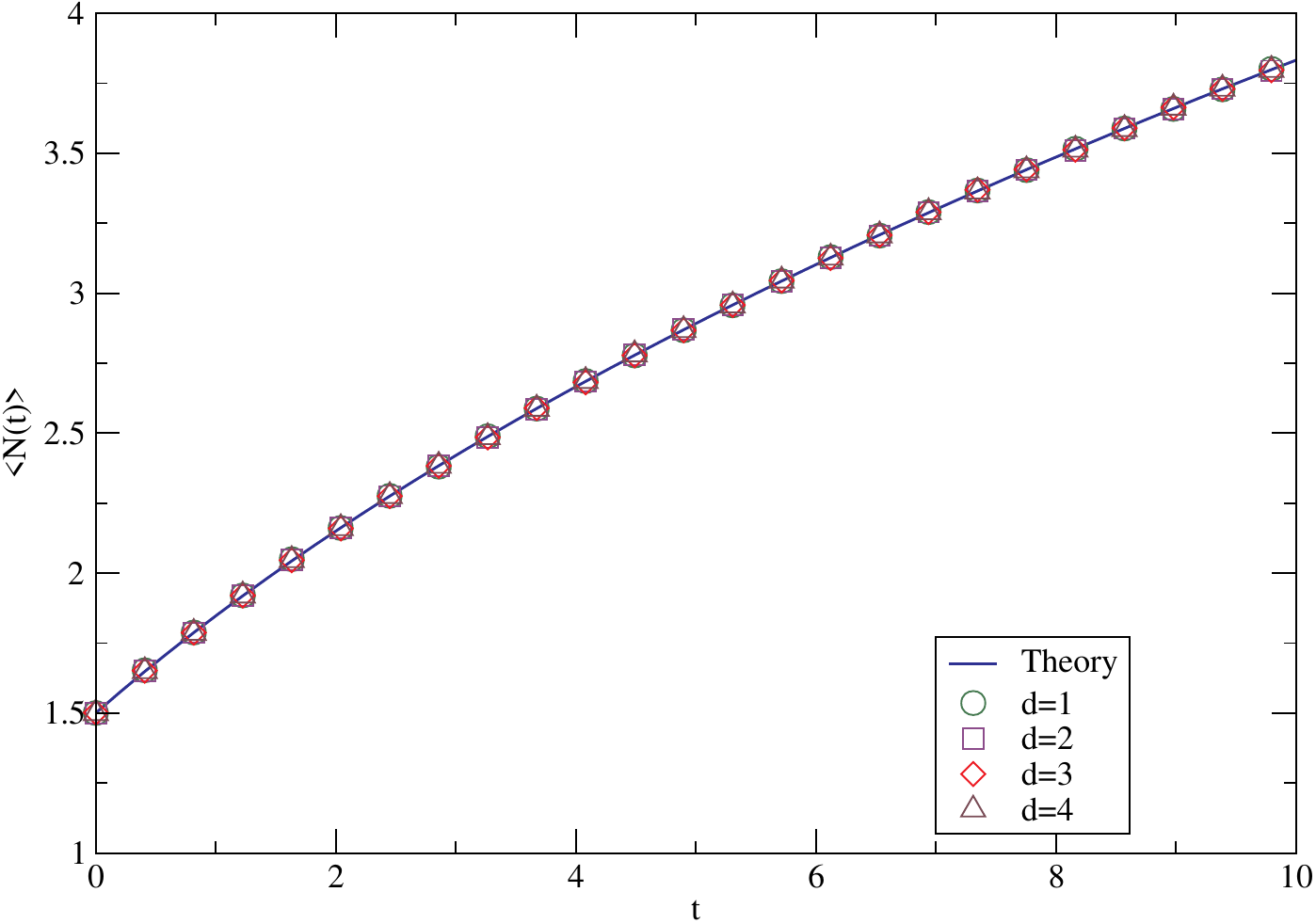}
\caption{Plot of the average number of records $\langle N(t) \rangle$ vs $t$. The solid line is given by the exact formula (\ref{av_record}) while the symbols represent numerical simulations in $d=1,2,3,4$ with $\gamma =1$ and $v_0 = 1$.}\label{fig_av_record}
\end{figure}

\begin{figure}[t]
\includegraphics[width = \linewidth]{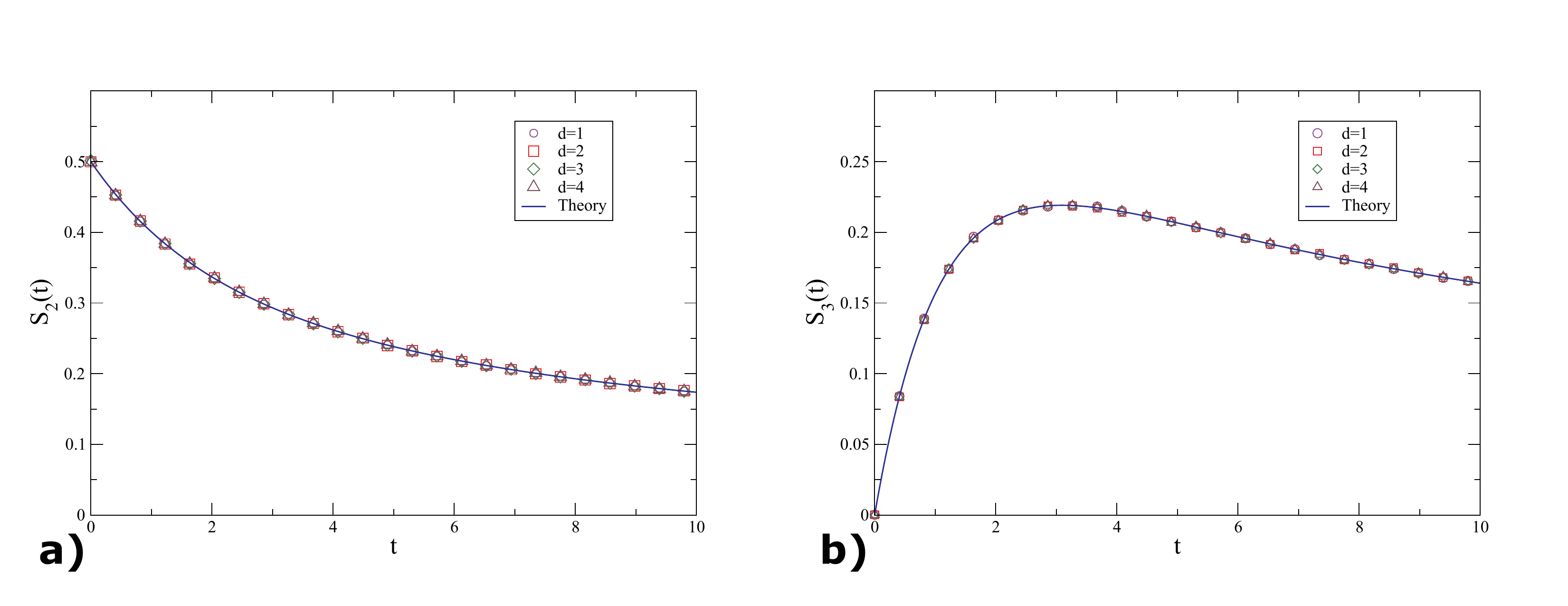}
\caption{Plot of $S_2(t)$ in (a) and $S_3(t)$ in (b) for different dimensions $d=1,2,3$ and $d=4$ (symbols correspond to numerical simulations) and an exponential distribution $p(\tau) = \gamma e^{-\gamma \tau}$ with $\gamma = 1$. The solid line corresponds to the exact results for $S_2(t) = S(t)$ in (\ref{S2}) and $S_3(t)$ in (\ref{S3}).}\label{Fig_s2_s3}
\end{figure}

Thanks to our mapping to the one-dimensional discrete-time random walk via Eq. (\ref{Pxn.1_supp}), we can use the known results for the record statistics of an $n$-step discrete-time random walk, whose steps are i.i.d. variables, each drawn from $\tilde p_s(x_i)$ which is continuous and symmetric, and given in Eq. (14) in the main text. It is well known that the probability $q_N(n)$ that a $n$-step random walk has exactly $N$ lower records is universal, i.e. independent of the distribution $\tilde p_s(x_i )$~\cite{Ziff_Satya}. In particular, its generating function with respect to $n$ is given by~\cite{Ziff_Satya} 
\begin{eqnarray} \label{GF_record_nber}
\sum_{n = N-1}^\infty q_N(n) z^n = \frac{(1-\sqrt{1-z})^{N-1}}{\sqrt{1-z}} \;.
\end{eqnarray}
The result in Eq. (\ref{Pxn.1_supp}) conveniently translates the results for any observable 
in the discrete-time $n$-step random walk problem to the RTP in continuous time $t$. The statistics
of records is one such observable. Therefore, combining Eq. (\ref{Pxn.1_supp}) gives (for $N \geq 2$)
\begin{eqnarray}\label{SN}
S_N(t) = \int \frac{ds}{2\pi\,i} e^{s\, t}\,
\frac{1}{\gamma}\, \sum_{n=N-1}^{\infty} \left(\frac{\gamma}{\gamma+s}\right)^n\, q_N(n) 
\end{eqnarray}
where the integral is over the Bromwich contour (imaginary axis in this case). Recall that, for $N=1$, $S_1(t) = S(t)$ where $S(t)$ is given in Eq. (1) of the main text. Using (\ref{GF_record_nber}) for $z = \gamma/(\gamma+s)$ in (\ref{SN}) one finds, for $N \geq 2$
\begin{eqnarray}\label{SN_2}
S_N(t) = \int \frac{ds}{2\pi\,i} e^{s\, t}\,
\frac{1}{\gamma}  \sqrt{\frac{\gamma + s}{s}}\left( 1 - \sqrt{\frac{s}{\gamma+s}}\right)^{N-1} \;.
\end{eqnarray}
The inverse Laplace transform on the right hand side of Eq. (\ref{SN_2}) can be performed explicitly for the first few values of $N$, yielding
\begin{eqnarray}\label{SN_3}
&&S_1(t) = S(t) = \frac{1}{2}e^{-\gamma t /2} \left( I_0 \left(\gamma t /2 \right) + I_1 \left( \gamma t /2\right) \right) \;, \label{S1}\\
&&S_2(t) = S(t) \;, \label{S2} \\
&&S_3(t) = e^{-{\gamma  t}/{2}} I_1\left({\gamma t}/{2}\right) \;. \label{S3}
\end{eqnarray}
The fact that $S_2(t)=S_1(t) = S(t)$ at all $t$ is quite remarkable and is far from obvious. These results for $N=2$ and $N=3$ are plotted in Fig. \ref{Fig_s2_s3} and one sees that $S_3(t)$ exhibits a maximum at some characteristic time $t_3^*$ (actually for all $N \geq 3$, $S_N(t)$ exhibits a maximum at some characteristic time $t^*_N$ which can be shown to grow linearly with $N$ for large $N$). It seems hard to evaluate explicitly $S_N(t)$ for higher values of $N$. One can however compute the generating function $\tilde S(z,t)$ of $S_N(t)$, i.e.
\begin{eqnarray}\label{GF_S}
\tilde S(z,t) = \sum_{N=1}^\infty z^N S_N(t) =  \int \frac{ds}{2\pi\,i} e^{s\, \gamma t}\, \left[ \frac{1+s}{s+ \frac{1-z}{z} \sqrt{s(1+s)}} - z\right] \;,
\end{eqnarray}
where we have made the change of variable $s \to s/\gamma$. Clearly $S_N(t)$ is universal, i.e. independent of the dimension $d$ and the speed distribution $W(v)$. From this expression, we can compute the average number of records $\langle N(t) \rangle$ up to time $t$ and we get, for all $t$ (see also Fig. \ref{fig_av_record})
\begin{eqnarray}\label{av_record}
\langle N(t) \rangle =  \frac{1}{2} e^{-\gamma t/2} \left((2 \gamma t+3) I_0\left(\frac{t}{2}\right)+(2 \gamma t+1)
   I_1\left(\frac{\gamma t}{2}\right)\right)  \;.
\end{eqnarray}
For large $t$, it grows like $\langle N(t) \rangle \approx 2\sqrt{\gamma t}/\sqrt{\pi}$. 

The Bromwich integral on the right hand side of Eq. (\ref{GF_S}) can be computed explicitly. Skipping details, we get
\begin{eqnarray}\label{GF_S2}
\tilde S(z,t) = \frac{z(1-z)}{1-2z} S(t)  - \frac{z^3}{1-2z} e^{-\frac{(1-z)^2}{1-2z} \gamma t} - \frac{z^3(1-z)}{(1-2z)^2} \gamma \int_0^{t} e^{-\frac{(1-z)^2}{1-2z}\gamma(t-t')} S(t') dt' \;,
 \end{eqnarray}
 where $S(t)$ is given in Eq. (\ref{S1}). By setting $z=1$ in Eq. (\ref{GF_S2}), we can check the normalization condition, i.e. $\sum_{N=1}^\infty S_N(t) = \tilde S(z=1,t) = 1$, for $t>0$. 
 We can also check, by expanding the generating function in (\ref{GF_S2}) in powers of $z$ up to order $z^3$, that we recover the results for $S_N(t)$ for $N=1,2,3$ in Eqs. (\ref{S1}-\ref{S3}). For generic $N$, we can check by expanding in powers of $z$ and performing the integral over $t'$ in Eq. (\ref{GF_S2}) that, for all $N$, $S_N(t)$ has the following structure, 
\begin{eqnarray}\label{GF_S3}
S_N(t) =  e^{-\gamma t/2} \left(P_{0,N}(\gamma t) I_0(\gamma t/2) + P_{1,N}(\gamma t) I_1(\gamma t/2)  \right) + e^{-\gamma t} Q_{N}(\gamma t) \;,
\end{eqnarray}
where $P_{0,N}(x), P_{1,N}(x)$ and $Q_N(x)$ are some polynomials.

One can also extract the asymptotic behaviors of $S_N(t)$ at small and large time $t$. At small time, from Eq. (\ref{SN}), one sees that the large $s$ behavior of the Laplace transform of $S_N(t)$ is $\sim \gamma^{N-2}\, q_{N-1}^{N}/s^{N-1}$, for $N \geq 2$. Using the known expression of $q_{N-1}^N = 2^{-N+1}$, from Ref. \cite{Ziff_Satya}, one obtains
\begin{eqnarray}\label{SN_small_t}
S_N(t) \sim \frac{(\gamma t)^{N-2}}{(N-2)!} \, q_{N-1}^N = \frac{1}{2^{N-1}(N-2)!} \, (\gamma t)^{N-2} \;.
\end{eqnarray} 
One sees explicitly that the small time behavior of $S_N(t)$ is dominated by trajectories where the RTP goes downwards at time $t$ and is
breaks a record at time $t$.  

The behavior of $S_N(t)$ for large time is easily obtained from the small $s$ of the Laplace transform in Eq. (\ref{SN_2}) and one finds, at leading order,
\begin{eqnarray} \label{SN_large_t}
S_N(t) \sim \frac{1}{\sqrt{\pi \gamma t}}  \;,
\end{eqnarray}
independently of $N$. This behavior indicates that $S_N(t)$ is dominated by the probability that, after breaking exactly $N$ lower records, the particle needs to stay above the value of the $N^{\rm th}$ record, which, for large $t$, coincides with the survival probability $S_1(t)\sim \frac{1}{\sqrt{\pi \gamma t}}$.